\documentclass[aps,pra,twocolumn,letterpaper,floatfix,superscriptaddress,10pt]{revtex4-2}

\usepackage{amsmath,amssymb}
\usepackage{mathtools}
\usepackage{graphicx}
\usepackage{bm}
\usepackage{enumitem}
\usepackage[pdftex,dvipsnames,usenames]{xcolor}
\usepackage[colorlinks=true,urlcolor=blue,citecolor=blue,linkcolor=blue]{hyperref}

\newcommand{\ket}[1]{\left|#1\right\rangle}
\newcommand{\bra}[1]{\left\langle #1\right|}
\DeclareMathOperator{\Tr}{Tr}
\newcommand{\normord}[1]{:\mathrel{#1}:}

\DeclareMathOperator{\Legendre}{P}

\begin{document}
	\title{Photocounting statistics of superconducting nanowire single-photon detectors}
	
	\author{V. A. Uzunova}
	\affiliation{Institute of Physics, NAS of Ukraine, Prospect Nauky 46, 03028 Kyiv, Ukraine}
	
	\author{A. A. Semenov}
	\affiliation{Bogolyubov Institute for Theoretical Physics, NAS of Ukraine, Vul. Metrologichna 14b, 03143 Kyiv, Ukraine}
	\affiliation{Institute of Physics, NAS of Ukraine, Prospect Nauky 46, 03028 Kyiv, Ukraine}
	\affiliation{Kyiv Academic University, Blvd. Vernadskogo  36, 03142  Kyiv, Ukraine}

	\date{\today}
	
	\begin{abstract}
		Superconducting nanowire single-photon detectors (SNSPDs) are efficient measurement devices used for counting single photons.
		The field of their applications covers experimental quantum-optical studies, optical quantum computing, quantum communication, and others.
		After registering a photon by such a detector, the next one cannot be registered during the dead time and after that this ability is smoothly restored.
		We have included this feature into the photodetection theory and introduced the corresponding photocounting formula.
		In the regime of continuous-wave detection, the photocounting statistics nonlinearly depends on the density operator due to a memory effect of previous measurement time windows.
		The considered examples demonstrate the strong influence of the relaxation process and the memory effect on the resulting photocounting statistics of the SNSPDs.
	\end{abstract}

	\maketitle
	
	
	\section{Introduction}
	\label{Sec:Intro}
	
	
	Fast and efficient single-photon detectors \cite{hadfield2009,eisaman11,natarajan12,you2020} are members of the fundamental building blocks in modern quantum technologies.
	The corresponding applications include optical universal \cite{Kok2007} and nonuniversal \cite{aaronson2013} quantum computers, quantum secure communication \cite{gisin07,Xu2020,Pirandola2020}, quantum sensing and metrology \cite{giovannetti11,degen17,polino2020}, etc.
	These detectors are also applied in a wider range of research and technologies such as the design of electronic devices \cite{somani01, Zhang03} and imaging problems in biological research and medicine  \cite{puszka2013,zysk2007,fujimoto1995}. 
	
	
	Superconducting nanowire single-photon detectors (SNSPD) \cite{semenov2001,goltsman2001,natarajan12,Zhang2020,you2020,zadeh2021} are highly promising measurement devices for numerous applications.  
	This is explained by their wide spectral sensitivity, good response speed, high detection efficiency, low dark count rate, and high temporal resolution.
	The detection process of the SNSPDs can be subdivided into a few stages \cite{zadeh2021}: 
	(1) an absorbed photon heats a small part of a nanowire in which a superconducting current flows;
	(2) this leads to forming a normal-conducting part of the nanowire resulting in a voltage change;
	(3) the detectors cannot register the next photons during the dead-time interval $\tau_\mathrm{d}$;  
	(4) the superconducting part of the nanowire and the ability to detect another photon are smoothly recovered during the relaxation time $\tau_\mathrm{r}$;
	(5) the detector returns to the initial state.
	It is worth noting that physical models \cite{goltsman2001,semenov2001,kerman2006,yang2007,SemenovAlexei2009,kerman2009,bulaevskii2012,zotova2012,engel2013,renema2014,engel2015,vodolazov2017,sunter2018,zhao2018,allmaras2019} describing the detection process in the SNSPDs are still developing.      
	
	The SNSPDs are applied in many fundamental quantum-optical experiments and in different implementations of quantum-information protocols \cite{you2020}.
	For example, they have been recently used for demonstration of quantum supremacy in the scheme of Gaussian boson sampling \cite{Zhong2020,zhong2021}.
	These detectors have been applied for implementations of quantum secure communication  \cite{hadfield2006,Takesue2007,Rosenberg2009,Liu2010,Sasaki2011,Wang2012,Tang2014,Tang2015,Takemoto2015,Valivarthi2015,Tang2016,Yin2016,boaron2018,Liu2019,Liu2019b,Wang2019,Minder2019,Fang2020,Chen2020,Wei2020} and quantum teleportation \cite{Sun2016,Valivarthi2016} protocols.
	The SNSPDs are key measurement elements for implementations of strong loophole-free tests of Bell inequalities \cite{Shalm2015}.
	They have also been applied as building blocks for array or time-multiplexing detectors in various quantum-optical experiments; see, e.g., \cite{Bohmann2016,Bohmann2018}.

	
	Photocounting is a prominent example of quantum measurements.
	Its outcomes are given by a number of clicks, which are commonly associated with the number of photons.
	According to Born's rule, the probability distribution to get $n$ clicks reads
		\begin{align}\label{Eq:PhotocountingEquationHS}
			P_n=\Tr\left(\hat{\rho}\,\hat{\Pi}_n\right),
		\end{align}	 
	where $\hat{\rho}$ is the density operator of the light mode and $\hat{\Pi}_n$ is the positive operator-valued measure (POVM) \cite{nielsen10} describing the measurement procedure.
	In the idealized scenario of photon-number resolving (PNR) detectors it is given by
		\begin{align}\label{Eq:POVM_PNR}
			\hat{\Pi}_n\equiv\hat{F}_n\left[\eta\right]=\normord{\frac{\left(\eta\hat{n}\right)^n}{n!}\exp\left(-\eta\hat{n}\right)},
		\end{align}
	cf. Refs.  \cite{mandel_book, kelley64}.
	Here $\hat{n}$ is the photon-number operator, $\eta\in[0,1]$ is the detection efficiency, and $\normord{\ldots}$ means the normal ordering. 
	In particular, this means that	$\hat{F}_n\left(1\right)=\ket{n}\bra{n}$ is the projector on the Fock state $\ket{n}$.	 
	
	In realistic scenarios, photon-number resolution is not ideal, which is described by the corresponding POVM.    
	For example, the POVM describing array \cite{paul1996,castelletto2007,schettini2007,blanchet08,Hlousek2019} and time-multiplexing \cite{achilles03,fitch03,rehacek03} detectors have been derived in Ref.~\cite{sperling12a}.
	The corresponding measurement procedures are based on spatial or temporal separation of modes and detecting each of them with on-off detectors. 
	
	Another scenario of realistic detection is based on counting the photocurrent pulses inside a measurement time window (MTW).
	The number of these pulses---referred to as clicks or photocounts---is associated with the number of photons.
	A problem is that each pulse is followed by a dead-time interval during which photons cannot be detected.
	A classical photocounting theory for this type of measurements has been developed in Refs.~\cite{ricciardi66,muller73,muller74,cantor75,teich78,vannucci78,Saleh_book,rapp2019,straka20}.

	
	In this paper we introduce a generalization of the photocounting theory to the scenario when the SNSPDs are used for counting pulses inside MTWs.
	The main difference of the SNSPDs from other detectors consists in the effect of the relaxation time.    
	The common problem of the SNSPDs and the detectors characterized by only dead time is that the corresponding time intervals from the last registered pulse may exceed the MTW.
	This results in changing the statistics of pulses for the next MTW.  
	
	We consider two detection scenarios associated with the SNSPDs.
	Firstly, the scenario of independent MTWs assumes darkening the detector at the end of each MTW.
	This protects statistics of pulses in the current MTW from influence of events in the previous ones.  
	Next, we consider the continuous-wave detection, which assumes no darkening at the ends of the MTWs.
	The corresponding photocounting statistics is affected by the previous MTWs---this influence is referred to as a memory effect. 
	It results in a nonlinear dependence of the photocounting statistics on the density operator.

	
	The rest of the paper is organized as follows.
	In Sec.~\ref{Sec:Preliminaries} we give a preliminary consideration of the photocounting formula in the Glauber-Sudarshan representation, which is used throughout the paper.
	The model of time-dependent efficiency, underlying the basis of our consideration of the SNSPDs, is discussed in Sec.~\ref{Sec:Preliminaries}.
	The POVM for the scenario of independent MTWs is derived in Sec.~\ref{Sec:IMTW}. 
	The scenario of continuous-wave detection involving the memory effects from previous MTWs is considered in Sec.~\ref{Sec:Continuos}.
	In Sec.~\ref{Sec:Examples} we apply the developed theory to deriving the photocounting statistics for typical quantum states.
	A technique for experimental reconstruction of the time-dependent efficiency is analyzed in Sec.~\ref{Sec:Xi}.
	A summary and concluding remarks are given in Sec.~\ref{Sec:Concl}.


\section{Preliminaries}
\label{Sec:Preliminaries}

	Photocounting formula (\ref{Eq:PhotocountingEquationHS}) can be conveniently rewritten as (see Refs.~\cite{cahill69,cahill69a})
		\begin{align}\label{Eq:PhotocountingEquationPS}	
			P_n=\int_\mathbb{C}d^2\alpha\, P(\alpha)\,\Pi_n(\alpha).
		\end{align}		 
	Here $P(\alpha)$ is the Glauber-Sudarshan P function \cite{glauber63c,sudarshan63} and
		\begin{align}
			\Pi_n(\alpha)=\bra{\alpha}\hat{\Pi}_n\ket{\alpha}
		\end{align}
	is the $Q$ symbols of the POVM defined as the average with the coherent state $\ket{\alpha}$.
	These symbols are interpreted as the probabilities to get $n$ clicks given the coherent state $\ket{\alpha}$.
	Utilizing the rule
		\begin{align}\label{Eq:Rule}
			\bra{\alpha}:\hat{f}\left(\hat{a},\hat{a}^\dag\right):\ket{\alpha}=f\left(\alpha,\alpha^\ast\right),
		\end{align}
	where $f\left(\alpha,\alpha^\ast\right)$ is an arbitrary function, one can reconstruct the normal-ordering operator form of the POVM from its $Q$ symbols.
	
	For the PNR detection, the $Q$ symbols of the POVM (\ref{Eq:POVM_PNR}) read
		\begin{align}\label{Eq:POVM_PNR_Q}
			\Pi_n(\alpha)\equiv F_n\left[\alpha;\eta\right]
			=\frac{\left(\eta|\alpha|^2\right)^n}{n!}\exp\left(-\eta|\alpha|^2\right).
		\end{align}	
	Here $F_n\left[\alpha;\eta\right]=\bra{\alpha}\hat{F}_n[\eta]\ket{\alpha}$ represents the $Q$ symbol of the operator $\hat{F}_n[\eta]$; cf. Eq.~(\ref{Eq:POVM_PNR}). 
	The first and second arguments of this function describe the dependence on the phase-space complex variable $\alpha$ and the detection efficiency $\eta$, respectively.
	This expression describes a well-known fact that photocounting statistics of the coherent states is given by the Poissonian distribution.
	Equation~(\ref{Eq:PhotocountingEquationPS}) has a form similar to the photocounting formula for classical electromagnetic fields \cite{mandel_book,Mandel_1964,Lamb1969}.
	In the classical theory both functions, $P(\alpha)$ and $\Pi_n(\alpha)$, are nonnegative.
	They play the roles of the probability density of the complex amplitude $\alpha$ and the classical response function of photocounts, respectively.
	
	For purposes of this work, it is also useful to remind a procedure of finding the POVM in the Fock-state basis, 
		\begin{align}\label{Eq:POVM_Fock}
			P_{n|m}=\bra{m}\hat{\Pi}_n\ket{m},
		\end{align}
	which can be interpreted as the probability distribution to get $n$ photocounts given $m$ photons.
	Since any kind of the photocounting measurements is phase insensitive, the nondiagonal POVM elements vanish, i.e., $\bra{m_1}\hat{\Pi}_n\ket{m_2}=0$ for $m_1\neq m_2$.
	The probabilities $P_{n|m}$ can also be considered as expansion coefficients of the POVM by the Fock states,
		\begin{align}\label{Eq:POVMexpan}
			\hat{\Pi}_n=\sum\limits_{m=0}^{+\infty}P_{n|m}\ket{m}\bra{m}.
		\end{align}  
	The same equation in terms of $Q$ symbols reads
		\begin{align}\label{Eq:CondProbTechn}
			\Pi_n(\alpha)\exp\left(|\alpha|^2\right)=\sum\limits_{m=0}^{+\infty}P_{n|m}
			\frac{|\alpha|^{2n}}{n!}.
		\end{align}
	Therefore, the POVM in the Fock-state basis can be obtained by expanding the left-hand side of this expression by $|\alpha|^{2n}/n!$.
	
	Another technique, which is used throughout the paper, is related to including realistic values of the detection efficiency and the dark-count rate.
	Let the POVM $\hat{\Pi}_n$ and its $Q$ symbols $\Pi_n(\alpha)$ describe the idealized scenario with the unit detection efficiency and with no dark counts.
	In order to include these issues in the description, one should replace 
		\begin{align}\label{Eq:ReplRealOp}
			\hat{n}\rightarrow\eta\hat{n}+\nu
		\end{align}
	and
		\begin{align}\label{Eq:ReplRealSymb}
			|\alpha|^2\rightarrow\eta|\alpha|^2+\nu
		\end{align}
	under the sign of normal ordering in the POVM and in the $Q$ symbols of the POVM, respectively; cf.~\cite{Pratt1969,Karp1970,Lee2005,Semenov2008}.
	Here $\eta$ and $\nu$ are the detection efficiency and the dark-count intensity, correspondingly.

	
	\section{Model of time-dependent efficiency}
	\label{Sec:DetectionModel}
	
	In this section we consider an idea, enabling us to study the effect of dead and relaxation times.
	For a correct formulation of our model, we should take into account two facts.
	Firstly, the probability to detect a photon is zero during the dead time $\tau_\mathrm{d}$ after the pulse.
	Secondly, the probability to detect a photon is smoothly recovered with the relaxation time $\tau_\mathrm{r}$, following the dead-time interval.  
	
	In the photodetection theory, the probability to detect a single photon is described by the detection efficiency.
	This means that we can consider the detection efficiency as a function of time $t$ passed after beginning of each pulse.
	This function is zero in the time-interval $[0,\tau_\mathrm{d}]$.
	After that the detection efficiency is smoothly recovered to its initial value.
	This implies that this function is given by
		\begin{align}\label{Eq:TDE}
			\xi(t)=\theta(t-\tau_\mathrm{d})\eta_\mathrm{r}(t-\tau_\mathrm{d}),
		\end{align}   
	where $\theta(t-\tau_\mathrm{d})$ is the Heaviside step-function and $\eta_\mathrm{r}(t)$ is the recovering efficiency.
	We chose the latter in the form
		\begin{align}\label{Eq:RDE}
			\eta_\mathrm{r}(t)=1-\exp\left({-\frac{t}{\tau_\mathrm{r}}}\right),
		\end{align} 
	which is used in our paper as a model of detector recovering.
	Although this model may be considered as an approximation---see, e.g., Refs.~\cite{Autebert2020,Caloz2017} for a more realistic description of the time-dependent detection efficiency $\xi(t)$---its advantage consists in possibilities of obtaining expressions suitable for analytical study.  
	Nevertheless, the main results of our paper are formulated in terms of an arbitrary function $\xi(t)$, which can also be reconstructed experimentally, as is discussed in Sec.~\ref{Sec:Xi} and in Ref.~\cite{Autebert2020}.
	In a more general context, the model given by the time-dependent efficiency $\xi(t)$ can be considered as a phenomenological description. 
	In principle, this model could be derived from a microscopic theory of quantum transitions in a superconducting nanowire.


\section{Photocounting with independent measurement time windows}
\label{Sec:IMTW}

\begin{figure}[htb]
		\centerline{\includegraphics[width = 1\linewidth]{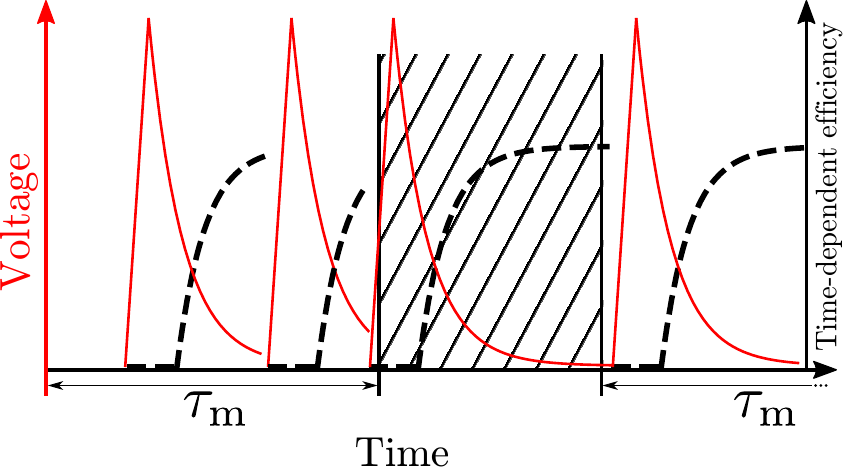}}
		\caption{\label{Fig:ITMW} Photocounting process with the SNSPDs in the scenario of independent MTWs is schematically depicted.
		The voltage pulses (solid lines) are counted during the MTWs duration of $\tau_\mathrm{m}$.
		According to Eq.~(\ref{Eq:TDE}), the time-dependent efficiency (dashed lines) is zero after registering each photon during the dead time $\tau_\mathrm{d}$, and after that it is smoothly recovered with the relaxation time $\tau_\mathrm{r}$.
		Detector input is darkened between the MTWs for the time interval sufficient for full recovering of the detector (hatched area).	
		}
	\end{figure}

	In this section we consider photocounting theory with the SNSPDs in the scenario of independent MTWs; see Fig.~\ref{Fig:ITMW}.
	The process of photocounting starts at the beginning of the time window of duration $\tau_\mathrm{m}$.  
	Each MTW is followed by a time interval of darkened detector-input sufficient for the detector to recover fully. 
	This eliminates the influence of the dead time and the relaxation process from the previous MTW on the measurement result in the current one.
	Alternatively, this technique can be implemented by a proper postselection of MTWs without darkening detector input.

\subsection{Positive operator-valued measure}
\label{Sec:POVM_IMTW}
	
	We start with consideration of the most general situation by assuming light being in a nonmonochromatic mode.
	For our purposes, such a mode can be characterized by an intensity function $I(t)$ normalized by the condition
		\begin{align}\label{Eq:NormI}
			\int\limits_0^{\tau_\mathrm{m}}d t I(t)=1.
		\end{align}
	The function $I(t)$ should be included explicitly in the POVM.
	In particular, this means that the detection efficiency in the time domain $[t,t+\Delta t]$ is given by
		\begin{align}
			\int\limits_{t}^{t+\Delta t}d t I(t)\xi(t-t_\mathrm{p}),
		\end{align}
	where $t_\mathrm{p}$ is the time moment at which the previous photon was detected.
	A particular example of the function $I(t)$ reads
		\begin{align}\label{Eq:IntMonochr}
			I(t)=\frac{1}{\tau_\mathrm{m}},
		\end{align} 
	which corresponds to a momochromatic light mode.
	For the sake of simplicity, we consider the unit detection efficiency and the zero dark-count rate.
	The corresponding generalization to realistic values of these parameters is straightforwardly obtained according to Eqs.~(\ref{Eq:ReplRealOp}) and (\ref{Eq:ReplRealSymb}). 
	
	Firstly, we note that the no-count element of the POVM is simply given by
		\begin{align}\label{Eq:No-countR}
			\hat{\Pi}_0=\hat{F}_0[1],
		\end{align}
	which properly describes absence of any pulses inside the MTW.
	Next, we derive the POVM element corresponding to the presence of a single pulse.
	For this purpose we consider the unnormalized probability density $\pi_1(t_1|\alpha)$ to get this pulse at the time moment $t_1$ given the coherent state $\ket{\alpha}$.
	The infinitesimal probability to get this click during the time interval $[t_1,t_1+d t_1]$ is obtained as the first expansion coefficient of the expression $1-F_0[\alpha;I(t_1)d t_1]$, cf. Eq.~(\ref{Eq:POVM_PNR_Q}), with respect to $d t_1$.
	This yields $|\alpha|^2I(t_1)d t_1$.
	It should be multiplied by the probabilities to get no-counts before and after the time moment $t_1$, which are given by
		\begin{align}\label{Eq:FirstPulse}
			F_0\left[\alpha;\int\limits_{0}^{t_1}d t I(t)\right]=\exp\left(-|\alpha|^2\int\limits_{0}^{t_1}d t I(t)\right)
		\end{align}	
	and 
		\begin{align}
			&F_0\left[\alpha;\int\limits_{t_1}^{\tau_\mathrm{m}}d t I(t)\xi(t-t_1)\right]\\
			&=\exp\left(-|\alpha|^2\int\limits_{t_1}^{\tau_\mathrm{d}}d t I(t)\xi(t-t_1)\right)\nonumber,
		\end{align}
	respectively.
	Therefore, the unnormalized probability density to get a single pulse at the time moment $t_1$ given the coherent state $\ket{\alpha}$ reads
		\begin{align}\label{Eq:POBM1}
			\pi_1\left(t_1|\alpha\right)={|\alpha|^{2}} I\left(t_1\right)\exp\left[-|\alpha|^2\Xi_1\left(\mathbf{t}\right)\right],
		\end{align}
	where
		\begin{align}\label{Eq:Xi_1}
			\Xi_1\left(t_1\right)=\int\limits_{0}^{t_1}d t I(t)
			+\int\limits_{t_1}^{\tau_\mathrm{m}}d t I(t)\xi(t-t_1).
		\end{align}	
	The $Q$ symbol of the corresponding POVM element,
		\begin{align}
			\Pi_1(\alpha)=\int\limits_0^{\tau_\mathrm{m}}d t_1 \pi_1\left(t_1|\alpha\right),
		\end{align}
	is obtained via integration with all possible values of $t_1$.
	
	As the last step, we derive the rest of the POVM elements, i.e. those for $n\geq2$. 
	Similar to the case of $n=1$, we derive the unnormalized probability density $\pi_{n}(\mathbf{t}|\alpha)$ to get pulses at the time moments $\mathbf{t}=(t_1,\ldots,t_n)$ given the coherent state $\ket{\alpha}$.
	This is composed of the following components:
		\begin{enumerate}[label=(\roman*)]
			\item The probability density to get a pulse at the time moment $t_1$ given by $|\alpha|^2I(t_1)$;
			\item The probability densities to get pulses at the time moments $t_i$ for $i=2\ldots n$  given by $|\alpha|^2I(t_i)\xi(t_i-t_{i-1})$;
			\item The probability to get no pulses from the time moment $t=0$ and up to the first pulse at the time moment $t_1$ given by Eq.~(\ref{Eq:FirstPulse});
			\item The probability to get no pulses in the time domains between the $i$th and $(i+1)$th pulses given by
				\begin{align}
					\exp\left(-|\alpha|^2\int\limits_{t_i}^{t_{i+1}}d tI(t)\xi(t-t_i)\right);
				\end{align}
			\item The probability to get no pulses in the time domain between the $n$th pulse and the time moment $t=\tau_\mathrm{m}$ given by 			
				\begin{align}
					\exp\left(-|\alpha|^2\int\limits_{t_n}^{\tau_\mathrm{m}}d tI(t)\xi(t-t_n)\right).
				\end{align}
		\end{enumerate}	
	Multiplying all these factors we arrive at the expression
		\begin{align}\label{Eq:POVM_DT_gen}
			\pi_n\left(\mathbf{t}|\alpha\right)={|\alpha|^{2n}} \mathcal{I}_n\left(\mathbf{t}\right)\exp\left[-|\alpha|^2\Xi_n\left(\mathbf{t}\right)\right],
		\end{align}  
	where
		\begin{align}\label{Eq:I}
			\mathcal{I}_n\left(\mathbf{t}\right)=I(t_1)\prod\limits_{i=2}^{n}I(t_i)\xi\left(t_i-t_{i-1}\right),
		\end{align}
	and
		\begin{align}\label{Eq:Xi}
			\Xi_n\left(\mathbf{t}\right)=\int\limits_{0}^{t_1}d t I(t)+\sum\limits_{i=1}^{n-1}\int\limits_{t_i}^{t_{i+1}}d t I(t)\xi(t-t_i)\nonumber\\
			+\int\limits_{t_n}^{\tau_\mathrm{m}}d t I(t)\xi(t-t_n).
		\end{align}
	Equation (\ref{Eq:POVM_DT_gen}) can be generalized to $n=1$ by setting $\Xi_1\left(t_1\right)$ in the form of Eq.~(\ref{Eq:Xi_1}) and $\mathcal{I}(t_1)=I(t_1)$.

	The $Q$ symbols of the POVM elements in the case of $n\geq1$ are given by
		\begin{align}\label{Eq:POVM_Q_gen}
			\Pi_n(\alpha)=\int_{T_n}d^n \mathbf{t}\, \pi_n\left(\mathbf{t}|\alpha\right),
		\end{align} 
	where integration is taken over the time-ordering domain $T_n$ such that $0\leq t_1\leq t_2\leq\ldots t_n\leq \tau_\mathrm{m}$.
	Employing property (\ref{Eq:Rule}), one obtains the general expression for the POVM elements
		\begin{align}\label{POVM_NM}
			\hat{\Pi}_n=:{\hat{n}^{n}}\int_{T_n}d^n\!\mathbf{t}\, \mathcal{I}_n\left(\mathbf{t}\right)\exp\left[-\hat{n}\Xi_n\left(\mathbf{t}\right)\right] :.
		\end{align}
	In the case of the time-dependent detection efficiency defined by Eq.~(\ref{Eq:TDE}), the maximal value of $n$ is restricted by the number $N+1$ or $N$, where $N=[\tau_\mathrm{m}/\tau_\mathrm{d}]$ is the number of whole dead-time intervals, fitting inside the MTW.
	The latter case is suitable  only if $N=\tau_\mathrm{m}/\tau_\mathrm{d}$.
	In order to generalize this POVM to the case of a nonunit detection efficiency $\eta$ and a nonzero dark-cont intesity $\nu$, one can use replacement described by Eqs.~(\ref{Eq:ReplRealOp}) and (\ref{Eq:ReplRealSymb}).
		
	For practical purposes, it is convenient to change the integration variables to $\boldsymbol{\tau}=\{\tau_0,\tau_1,\tau_2,\ldots\tau_n\}$,
		\begin{align}
			&\tau_0=t_1,\\
			&\tau_i=t_{i+1}-t_i,\\
			&\tau_n=\tau_\mathrm{m}-t_n. \label{Eq:Tau_n}
		\end{align} 
	An inverse relation,
		\begin{align}\label{Eq:ChangeVar}
			t_i=\sum\limits_{j=0}^{i-1}\tau_i
		\end{align}
	can be substituted in Eqs.~(\ref{Eq:I}) and (\ref{Eq:Xi}).
	The new variables correspond to the time intervals between neighboring pulses.
	Herewith, $\tau_0$ and $\tau_n$ are the intervals from the beginning of the MTW up to the first pulse and from the last pulse up to the end of the MTW, respectively.
	These nonnegative variables obey the constraint
		\begin{align}
			\sum\limits_{i=0}^{n}\tau_i=\tau_\mathrm{m},
		\end{align}
	which defines an $n$-dimensional simplex of $(n{+}1)$-dimensional space.

	\subsection{Regular and irregular parts of the POVM}

	If we consider the time-dependent efficiency in the form of Eq.~(\ref{Eq:TDE}), the POVM can further be specified as
		\begin{align}\label{Eq:POVM-r-i}
			\hat{\Pi}_n=\hat{\Pi}^{\mathrm{(r)}}_n+\hat{\Pi}^{\mathrm{(i)}}_n,
		\end{align}
	where $\hat{\Pi}^{\mathrm{(r)}}_n$ and $\hat{\Pi}^{\mathrm{(i)}}_n$ are referred to as regular and irregular parts, respectively.
	The regular part describes the situation with all $n$ dead-time intervals fitted inside the MTW.
	The irregular part describes the situation, for which the last dead-time interval exceeds the MTW.    
	This representation is made by splitting the integral for $\tau_n$ in two parts.
	
	Let us derive analytical expressions for regular and irregular parts of the POVM.
	Herein the presence of the Heaviside theta-function in Eq.~(\ref{Eq:TDE}) will be accounted in the integration domain. 
	In this case, the regular part for $n=0\ldots N$ is given by
		\begin{align}\label{Eq:POVM-r}
			&\hat{\Pi}^{\mathrm{(r)}}_n=\\
			&:{\hat{n}^n}\int\limits_{\tau_\mathrm{d}}^{\tau_\mathrm{m}-(n-1)\tau_\mathrm{d}}d\tau_n\int_{\Omega^{n}}d^{n-1}\!\boldsymbol{\tau} \mathcal{J}_n(\boldsymbol{\tau})e^{-{\hat{n}}{\Xi}^{\mathrm{(r)}}_n(\boldsymbol{\tau})}:.\nonumber
		\end{align}
	For $n=N+1$ this part vanishes.
	Here 
		\begin{align}
			\mathcal{J}_1(\boldsymbol{\tau})=I(\tau_0),
		\end{align}
		\begin{align}
			\mathcal{J}_n(\boldsymbol{\tau})=I(\tau_0)\prod\limits_{i=2}^{n}I\left(\sum\limits_{j=0}^{i-1}\tau_i\right)\eta_\mathrm{r}(\tau_{i-1}-\tau_\mathrm{d})
		\end{align}
	for $n\geq2$,
		\begin{align}
			{\Xi}^{\mathrm{(r)}}_n&(\boldsymbol{\tau})=
			\int\limits_{0}^{\tau_0}d tI(t)\\
			&+\sum\limits_{j=1}^{n}\int\limits_{\tau_\mathrm{d}}^{\tau_j}d tI\left(\sum\limits_{k=0}^{j-1}\tau_k+t\right)\eta_{\mathrm{r}}(t-\tau_\mathrm{d}).\nonumber
		\end{align}
	The integration domain $\Omega_n$ is defined as
		\begin{align}
			\int_{\Omega^{n}}d^{n-1}\!\boldsymbol{\tau}\ldots=\int\limits_{\tau_\mathrm{d}}^{\Delta_{n-1}}d\tau_{n-1}\ldots
			\int\limits_{\tau_\mathrm{d}}^{\delta_1}d\tau_{1}\ldots,
		\end{align}
	where
		\begin{align}
			\delta_i=\tau_\mathrm{m}-(i-1)\tau_\mathrm{d}-\sum_{j=i+1}^n\tau_{j}.
		\end{align}
	As mentioned, this integration  is accounted for by the zero-value domains of the Heaviside step-functions in Eq.~(\ref{Eq:TDE}).
	
	The irregular part of the POVM reads
		\begin{align}\label{Eq:POVM-i}
			&\hat{\Pi}^{\mathrm{(i)}}_n=\\
			&:{\hat{n}^n}\int\limits_{0}^{\tau_\mathrm{d}}d\tau_n\int_{\Omega^{n}}d^{n-1}\!\boldsymbol{\tau} \mathcal{J}_n(\boldsymbol{\tau})e^{-{\hat{n}}{\Xi}^{\mathrm{(i)}}_n(\boldsymbol{\tau})}:,\nonumber
		\end{align}		
	 for $n=0\ldots N$ and
	 	\begin{align}\label{Eq:POVM-i-Np1}
			 \hat{\Pi}^{\mathrm{(i)}}_{N+1}&=:{\hat{n}^{N+1}}\int\limits_{0}^{\tau_\mathrm{m}-N\tau_\mathrm{d}}d\tau_{N+1}\\
			 &\times\int_{\Omega^{N+1}}d^{N}\!\boldsymbol{\tau} \mathcal{J}_{N+1}(\boldsymbol{\tau})e^{-{\hat{n}}{\Xi}^{\mathrm{(i)}}_{N+1}(\boldsymbol{\tau})}:.\nonumber
		 \end{align}	
	 Here
		\begin{align}
			{\Xi}^{\mathrm{(i)}}_n&(\boldsymbol{\tau})=
			\int\limits_{0}^{\tau_0}d tI(t)\\
			&+\sum\limits_{j=1}^{n-1}\int\limits_{\tau_\mathrm{d}}^{\tau_j}d tI\left(\sum\limits_{k=0}^{j-1}\tau_k+t\right)\eta_{\mathrm{r}}(t-\tau_\mathrm{d}).\nonumber
		\end{align}
	The integration domain $\Omega_n$ and the function $\mathcal{J}_n(\boldsymbol{\tau})$ are the same as for the regular part. 	
	
	Consider the case of a monochromatic mode [cf. Eq.~(\ref{Eq:IntMonochr})] and the recovering detector efficiency in the form of Eq.~(\ref{Eq:RDE}).
	This gives a possibility to present the functions $\mathcal{J}_n(\boldsymbol{\tau})$, ${\Xi}^{\mathrm{(r)}}_n(\boldsymbol{\tau})$, and ${\Xi}^{\mathrm{(i)}}_n(\boldsymbol{\tau})$ in the explicit form,
		\begin{align}
			\mathcal{J}_n(\boldsymbol{\tau})=\frac{1}{\tau_\mathrm{m}^n}\prod_{j=1}^{n-1}\eta_\mathrm{r}(\tau_j-\tau_\mathrm{d})
		\end{align}
	for $n\geq2$,
		\begin{align}
			{\Xi}^{\mathrm{(r)}}_n(\boldsymbol{\tau})=
			\eta_n-\frac{\tau_\mathrm{r}}{\tau_\mathrm{m}}\sum_{j=1}^{n}\eta_\mathrm{r}(\tau_j-\tau_\mathrm{d}),
		\end{align}
		\begin{align}
			{\Xi}^{\mathrm{(i)}}_n(\boldsymbol{\tau})=
		\eta_{n-1}-\frac{\tau_n}{\tau_\mathrm{m}}-\frac{\tau_\mathrm{r}}{\tau_\mathrm{m}}\sum_{j=1}^{n-1}\eta_\mathrm{r}(\tau_j-\tau_\mathrm{d}).
		\end{align}
	Here 
		\begin{align}\label{Eq:ADE}
			\eta_n=\frac{\tau_\mathrm{m}-n\tau_\mathrm{d}}{\tau_\mathrm{m}}
		\end{align}
	is the adjusting detection efficiency describing the time free from the dead-time interval as a ratio to the duration of the MTW.
	
	An important example corresponds to the case of zero relaxation time, $\tau_\mathrm{r}=0$, such that $\eta_\mathrm{r}=1$.
	This situation takes place for many types of detectors, such as avalanche photodiodes and photomultiplier tubes. 
	The regular part of the POVM of such detectors for $n=0\ldots N$ is given by
		\begin{align}\label{Eq:POVM_simple0}
			 \hat{\Pi}_n^\mathrm{(r)}=
			 \hat{F}_n\big[\eta_n\big],
		\end{align}
	where $\eta_n$ is the adjusting detection efficiency and $\hat{F}_n\big[\eta\big]$ is the POVM of the PNR detectors, cf. Eqs.~(\ref{Eq:ADE}) and (\ref{Eq:POVM_PNR}), respectively.
	For $n=N+1$ this part vanishes.
	The irregular part of the corresponding POVM reads
		\begin{align}\label{Eq:POVM_simple}
			\hat{\Pi}_n^\mathrm{(i)}=\sum\limits_{k=0}^{n-1}\hat{F}_k\left[\eta_n\right]-
			\sum\limits_{k=0}^{n-1}\hat{F}_k\left[\eta_{n-1}\right]
		\end{align}
	for $n=0\ldots N$ and 
		\begin{align}\label{Eq:POVM_simpleNp1}
			\hat{\Pi}_{N+1}^\mathrm{(i)}=1-
			\sum\limits_{k=0}^{N}\hat{F}_k\left[\eta_{N}\right].
		\end{align}
	 for $n=N+1$.
	 Combining both parts of the POVM in Eq.~(\ref{Eq:POVM-r-i}), we arrive at the POVM, the $Q$ symbols of which correspond to the classical photodetection theory with the dead time \cite{ricciardi66,muller73,muller74,cantor75,teich78,vannucci78,Saleh_book,rapp2019,straka20}.
		
	We remind that all expressions presented here are given for the unit detection efficiency $\eta$ and the zero dark-count intensity $\nu$.
	These equations can be simply rewritten to the case of realistic values of these quantities.
	For this purpose, one should use the replacement described by Eqs.~(\ref{Eq:ReplRealOp}) and (\ref{Eq:ReplRealSymb}).

\subsection{Photocounting statistics vs photon-number statistics}

	As follows from the previous consideration, the photocounting statistics (statistics of pulse numbers) may significantly differ from the photon-number statistics.
	Averaging both sides of Eq.~(\ref{Eq:POVMexpan}) with the density operator $\hat{\rho}$, we arrive at the linear expression,
		\begin{align}\label{Eq:PulseViaPhot}
			P_n=\sum\limits_{m=0}^{+\infty}P_{n|m}\mathcal{P}_m,
		\end{align}
	connecting the photocounting distribution $P_n$ with the photon-number distribution $\mathcal{P}_m=\bra{m}\hat{\rho}\ket{m}$ in the case of $\eta=1$ and $\nu=0$. 
	A similar expression in the case of array detectors is applied for reconstructions of $\mathcal{P}_m$ from $P_n$ via regularization of an ill-posed problem \cite{Hlousek2019}.
	For realistic values of $\eta$ and $\nu$, one can use $\mathcal{P}_m=\Tr\left(\hat{\rho}\hat{F}_m[\eta;\nu]\right)$, where $\hat{F}_m[\eta;\nu]$ is obtained from $\hat{F}_m[1]$ [cf. Eq.~(\ref{Eq:POVM_PNR})] by the replacement (\ref{Eq:ReplRealOp}).
	This procedure does not result in changing the conditional probabilities $P_{n|m}$; cf. Ref.~\cite{Kovalenko2018}.
	However, it enables one to consider the effects caused solely by the realistic resolution of photon numbers.

	We apply the technique described by Eq.~(\ref{Eq:CondProbTechn}) to the POVM (\ref{POVM_NM}). 
	This yields the expression for the probability to get $n$ pulses given $m$ photons,
		\begin{align}\label{Eq:POVM-Fock}
			P_{n|m}=\frac{m!}{(m-n)!}\int_{T_n}d^n\!\mathbf{t}\, \mathcal{I}_n\left(\mathbf{t}\right)\left[1-\Xi_n\left(\mathbf{t}\right)\right]^{m-n}
		\end{align}
	for $m\geq n$ and 
		\begin{align}\label{Eq:POVM-Fock0}
			P_{n|m}=0
		\end{align}
	for $m< n$.
	The latter means that we cannot get more pulses than photons at the detector input. 
	Integration in Eq.~(\ref{Eq:POVM-Fock}) is performed in the time-ordering domain $T_n$.

	Applying the same method to Eq.~(\ref{Eq:POVM-r}) and Eqs.~(\ref{Eq:POVM-i}) and (\ref{Eq:POVM-i-Np1}), we can subdivide the conditional probability $P_{n|m}$ on regular and irregular parts, respectively.
	Specifically, the regular part is given by
		\begin{align}\label{Eq:POVM-Fock-r}
			&P_{n|m}^{\mathrm{(r)}}=\frac{m!}{(m-n)!}\\
			&\times\int\limits_{\tau_\mathrm{d}}^{\tau_\mathrm{m}-(n-1)\tau_\mathrm{d}}d\tau_n\int_{\Omega^{n}}d^{n-1}\!\boldsymbol{\tau} \mathcal{J}_n(\boldsymbol{\tau})\left[1-{\Xi}^{\mathrm{(r)}}_n(\boldsymbol{\tau})\right]^{m-n}\nonumber
		\end{align}		
	for $n=0\ldots N$.
	The irregular part reads
		\begin{align}\label{Eq:POVM-Fock-i}
			&P_{n|m}^{\mathrm{(i)}}=\frac{m!}{(m-n)!}\\
			&\times\int\limits_{0}^{\tau_\mathrm{d}}d\tau_n\int_{\Omega^{n}}d^{n-1}\!\boldsymbol{\tau} \mathcal{J}_n(\boldsymbol{\tau})\left[1-{\Xi}^{\mathrm{(i)}}_n(\boldsymbol{\tau})\right]^{m-n}\nonumber
		\end{align}				
	for $n=0\ldots N$ and
		\begin{align}\label{Eq:POVM-Fock-i-Np1}
			&P_{N+1|m}^{\mathrm{(i)}}=\frac{m!}{(m-N-1)!}\int\limits_{0}^{\tau_\mathrm{m}-N\tau_\mathrm{d}}d\tau_{N+1}\\
			&\times\int_{\Omega^{n}}d^{N}\!\boldsymbol{\tau} \mathcal{J}_{N+1}(\boldsymbol{\tau})\left[1-{\Xi}^{\mathrm{(i)}}_{N+1}(\boldsymbol{\tau})\right]^{m-N-1}\nonumber
		\end{align}		
	for $n=N+1$.
	Both parts vanish for $m<n$.
	
	Let us consider a special model: the monochromatic mode given by Eq.~(\ref{Eq:IntMonochr}) and the recovering detector efficiency in the form of Eq.~(\ref{Eq:RDE}).
	In this case, integrals in Eqs.~(\ref{Eq:POVM-Fock-r}), (\ref{Eq:POVM-Fock-i}), and (\ref{Eq:POVM-Fock-i-Np1}) can be evaluated analytically for each values of $n$ and $m$ although the general equations have a complex structure.
	The corresponding expressions for $P^\mathrm{(r)}_{n|m}$ and  $P^\mathrm{(i)}_{n|m}$ can be directly used in Eq.~(\ref{Eq:PulseViaPhot}) for deriving the photocounting statistics from the photon-number statistics.
		
	In the important case of $n=m$, the probabilities $P_{n|n}$ are reduced to a simple analytical form,
		\begin{align}\label{Eq:CorrCoeff}
			P_{n|n}=
			\frac{\tau_\mathrm{r}^{n}}{\tau_\mathrm{m}^{n}}\left[\sum_{l=0}^{n}a^l_{n-1}(-1)^{n-l}\frac{n!(2n-2-l)!}{l!(n-l)!(n-2)!}\right.
			\\
			+\left.(-1)^{n-1}e^{-a_{n-1}}\sum_{l=0}^{n-2}a^l_{n-1}\frac{(2n-2-l)!}{l!(n-l-2)!}\right],\nonumber
		\end{align}
	where
		\begin{align}
			a_n=\frac{\tau_\mathrm{m}-n\tau_\mathrm{d}}{\tau_\mathrm{r}}.
		\end{align}
	The quantity (\ref{Eq:CorrCoeff}) characterizes the ability of detectors to distinguish between photon numbers.
	For the PNR detectors, i.e. for $\tau_\mathrm{d}=\tau_\mathrm{r}=0$, it takes the unit value.
	It tends to zero for the detectors, for which the click number is never equal to the number of detected photons.

\section{Continuous-wave detection}
\label{Sec:Continuos}

\begin{figure}[htb]
			\includegraphics[width = 1\linewidth]{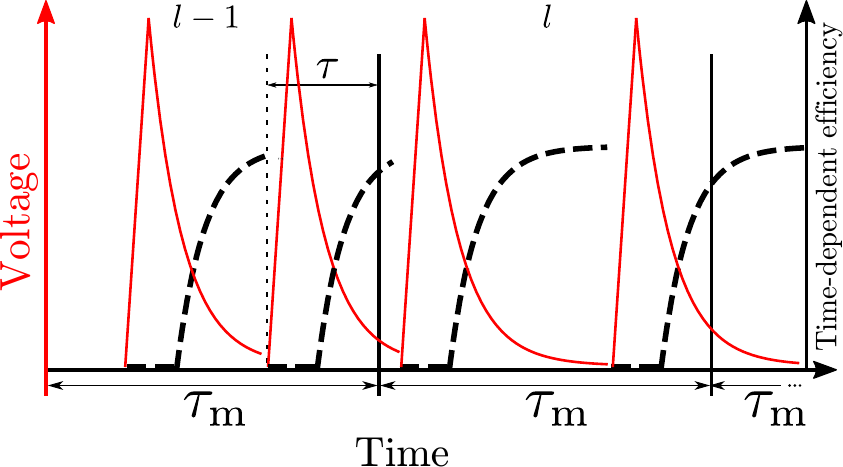}
			\caption{\label{Fig:CWD} Photocounting with the technique of continuous-wave detection is schematically depicted.
			The voltage pulses (solid lines) are counted inside the MTW.
			The time-dependent efficiency $\xi(t)$ is shown by dashed lines. 
			The time $\tau$ between the last pulse in the $(l-1)$th MTW and its end is shown.}
		\end{figure}

	A typical photocounting technique of the continuous-wave detection assumes no interruptions between the MTWs; see Fig.~\ref{Fig:CWD}.
	In such a scenario, the detector may not be recovered after the last pulse from the previous MTW.
	This affects on the probability of events in the current MTW.   
	Therefore, photocounting statistics depends on quantum states in previous MTWs.
		
	Let us numerate MTWs in the order as they appear in time.
	The $Q$ symbols of the POVM in the $l$th MTW, $\Lambda_n\left(\boldsymbol{\alpha}^l\right)$, depend on the amplitudes $\boldsymbol{\alpha}^l=\begin{pmatrix} \alpha_1, \ldots \alpha_l\end{pmatrix}$ in the given and all previous MTWs.
	Here and in the following consideration, we use the upper indices of bold symbols in order to designate the number of entries in the corresponding sets of numbers.
	Therefore, the photocounting formula in the P representation is given by  
		\begin{align}\label{Eq:PhCountEq_Full_P}
			P_n=\int_{\mathbb{C}^l}d^{2l}\boldsymbol{\alpha}^l\, P(\alpha_l) P(\alpha_{l-1})\ldots P(\alpha_1)\,\Lambda_n\big(\boldsymbol{\alpha}^{l}\big).
		\end{align}
	This formula in the operator form reads
		\begin{align}\label{Eq:PhotocountingEquationGeneral}
			P_n=\Tr\left(\hat{\rho}^{\otimes l}\,\hat{\Lambda}_n\right).
		\end{align}
	The most significant difference of these relations from Eqs.~(\ref{Eq:PhotocountingEquationHS}) and (\ref{Eq:PhotocountingEquationPS}) consists in nonlinear dependence of the photocounting distribution on the density operator $\hat{\rho}$.
	In this section we will derive the POVM $\hat{\Lambda}_n$ for the scenario of the continuous-wave detection in an explicit form and consider the memory effect of the previous MTWs on the photocounting distribution in the current one.

\subsection{Events dependent on previous measurements}

	Before we start with deriving the general expression for the POVM $\hat{\Lambda}_n$ in the scenario of continuous-wave detection, we consider an important part needed for this consideration.
	Let us assume that the last pulse in the $(l-1)$th MTW occurs at the time moment $\tau_n=\tau$ [cf. Eq.~(\ref{Eq:Tau_n})] before its end; see Fig.~\ref{Fig:CWD}.
	The POVM in the $l$th MTW should now depend on this time.

	The no-count element of the POVM in the considered case differs from Eq.~(\ref{Eq:No-countR}) and is given by
		\begin{align}
			\hat{\Pi}_0(\tau)=:\exp\left[-\Xi_0(\tau)\hat{n}\right]:,
		\end{align}
	where 
		\begin{align}\label{Eq:Xi0Tau}
			\Xi_0(\tau)=\int\limits_{0}^{\tau_\mathrm{m}}d t I(t)\xi(t+\tau)
		\end{align}
	corresponds to the no-count efficiency.
	Derivation of all other elements of the POVM resembles the points listed in Sec.~\ref{Sec:POVM_IMTW} but with two modifications.
	Firstly, the unnormalized probability density to get the first pulse at the time moment $t_1$ is now given by $|\alpha|^2I(t_1)\xi(\tau+t_1)$.
	Secondly, the probability to get no pulses before the time moment $t_1$ is modified to the form  
		\begin{align}\label{Eq:FirstPulseTau}
			F_0&\left[\alpha;\int\limits_{0}^{t_1}d t I(t)\xi(\tau+t)\right]\\
			&=\exp\left(-|\alpha|^2\int\limits_{0}^{t_1}d t I(t)\xi(\tau+t)\right).\nonumber
		\end{align}
	Other points of the derivation are not changed. 	
	
	The $Q$ symbols of the POVM conditioned by the time $\tau$ for $n\geq 1$ are given by
		\begin{align}\label{POVM_Symb_Tau}
			\Pi_n(\alpha;\tau)=\int_{T_n}d^n\!\mathbf{t}\,\pi_n\left(\mathbf{t}|\alpha;\tau\right).
		\end{align} 
	Here $\pi_n\left(\mathbf{t}|\alpha;\tau\right)$ is the unnormalized probability density to get pulses at the time moments $\mathbf{t}$ given the coherent states $\ket{\alpha}$ and the time $\tau$ between the last pulse in the previous MTW and its end.
	This conditional probability density reads
		\begin{align}\label{POVM_Symb_Tau_Dens}
			\pi_n\left(\mathbf{t}|\alpha;\tau\right)={|\alpha|^{2n}} \mathcal{I}_n\left(\mathbf{t};\tau\right)\exp\left[-|\alpha|^2\Xi_n\left(\mathbf{t};\tau\right)\right].
		\end{align}
	Here
		\begin{align}\label{Eq:ITau}
			\mathcal{I}_n\left(\mathbf{t};\tau\right)=\mathcal{I}_n\left(\mathbf{t}\right)\xi(\tau+t_1)
		\end{align}
	[cf.~Eq.~(\ref{Eq:I}) for $\mathcal{I}_n\left(\mathbf{t}\right)$],
		\begin{align}\label{Eq:XiTau}
			&\Xi_n\left(\mathbf{t};\tau\right)=\int\limits_{0}^{t_1}d t I(t)\xi(\tau+t)\nonumber\\
			&+\sum\limits_{i=1}^{n-1}\int\limits_{t_i}^{t_{i+1}}d t I(t)\xi(t-t_i)
			+\int\limits_{t_n}^{\tau_\mathrm{m}}d t I(t)\xi(t-t_n)
		\end{align}
	for $n\geq 2$ and 
		\begin{align}\label{Eq:Xi1Tau}
			\Xi_1\left(t_1;\tau\right)=\int\limits_{0}^{t_1}d t I(t)\xi(\tau+t)
			+\int\limits_{t_1}^{\tau_\mathrm{m}}d t I(t)\xi(t-t_1)
		\end{align}
	for $n=1$.
	
	To summarize this part, we note that the POVM conditioned by the time $\tau$ is given by
		\begin{align}\label{POVM_Tau}
			\hat{\Pi}_n(\tau)=:{\hat{n}^{n}}\int_{T_n}d^n\!\mathbf{t}\, \mathcal{I}_n\left(\mathbf{t};\tau\right)\exp\left[-\hat{n}\Xi_n\left(\mathbf{t};\tau\right)\right] :.
		\end{align}
	Here $\mathcal{I}_n\left(\mathbf{t};\tau\right)$ is given by Eq.~(\ref{Eq:ITau}) for $n\geq 1$ and $I_0=1$.
	Similarly, $\Xi_n\left(\mathbf{t};\tau\right)$ is given by Eqs.~(\ref{Eq:XiTau}), (\ref{Eq:Xi1Tau}), and (\ref{Eq:Xi0Tau}) for $n\geq 2$, $n=1$, and $n=0$, respectively.
	
\subsection{POVM in the Markovian approximation}

	In this subsection we derive an expression for the POVM in the scenario of the continuous-wave detection considering the measurement in the $l$th MTW; cf. Eqs.~(\ref{Eq:PhCountEq_Full_P}) and (\ref{Eq:PhotocountingEquationGeneral}).
	In the next sections we will show that the dependence on the number $l$ is negligible for $l\gg 1$, and its actual value is not important.
	However, for the sake of consistency, this number should be explicitly included in our consideration.
	We also assume that the ratio of dead and relaxation times to the duration of the MTW is small, 
		\begin{align}\label{Eq:CondTime}
			\frac{\tau_\mathrm{d}+\tau_\mathrm{r}}{\tau_\mathrm{m}}\ll 1.
		\end{align}  
	This yields for the time-dependent efficiency 
		\begin{align}\label{Eq:CondXiMA}
			\xi(t\geq \tau_\mathrm{m})\approx 1,
		\end{align}
	i.e. if no pulses are registered in the $(l-1)$th MTW, then the pulses in the former MTWs do not affect the statistics in the $l$th MTW.
	This is the essence of the Markovian approximation considered here. 
	 
	Similar to Sec.~\ref{Sec:IMTW}, we consider the $Q$ symbols of the POVM as the probabilities to get $n$ pulses given the coherent state $\ket{\alpha}$.
	Let $\rho^{(k)}\left(\tau|\boldsymbol{\alpha}^{k}\right)$ be the probability density for the time-interval [between the last pulse and the end of the MTW] $\tau$ in the $k$th MTW given the coherent amplitudes $\boldsymbol{\alpha}^{k}$ in this and in all previous MTWs.
	This function is normalized by the condition
		\begin{align}\label{Eq:Normalization}
			\Lambda^{(k)}_0\left(\boldsymbol{\alpha}^{k}\right)+\int\limits_{0}^{\tau_\mathrm{m}}d \tau \rho^{(k)}\left(\tau|\boldsymbol{\alpha}^{k}\right)=1,
		\end{align}
	where $\Lambda^{(k)}_0\left(\boldsymbol{\alpha}^{k}\right)$ is the $0$th element of the $Q$ symbols of the POVM, i.e. the probability to get no pulses in the $k$th MTW given the coherent amplitudes $\boldsymbol{\alpha}^{k}$.
	The form of these functions will be specified latter.
	
	Two kinds of the outcomes in the $(l-1)$th MTW affect on the statistics in the $l$th one:
		\begin{enumerate}[label=(\roman*)]
			\item No pulses in the $(l-1)$th MTW with the probability $\Lambda^{(l-1)}_0\left(\boldsymbol{\alpha}^{l-1}\right)$ occur.
			In this case, the unnormalized probability density to get pulses at the time moments $\textbf{t}^n$ of the $l$th MTW is $\pi_n(\textbf{t}^n|\alpha_l)$. 
			
			\item Pulses in the $(l-1)$th MTW occur such that the probability density for the time-interval $\tau$ is  $\rho^{(l-1)}\left(\tau|\boldsymbol{\alpha}^{l-1}\right)$.
			In this case, the unnormalized probability density to get pulses at the time moments $\textbf{t}^n$ of the $l$th MTW is $\pi_n(\textbf{t}^n|\alpha_l;\tau)$. 
		\end{enumerate}
	Employing the law of total probability, we get the unnormalized probability density to get pulses at the time moments $\textbf{t}^n$ of the $l$th MTW given the coherent amplitudes $\boldsymbol{\alpha}^l$ in current and all previous MTWs in the form
		\begin{align}\label{Eq:POVM_Dens_Gen}
			\lambda^{(l)}_n(\textbf{t}^n&|\boldsymbol{\alpha}^l)=\pi_n(\textbf{t}^n|\alpha_l)\Lambda^{(l-1)}_0\left(\boldsymbol{\alpha}^{l-1}\right)\\
			&+\int\limits_{0}^{\tau_\mathrm{m}}d \tau \pi_n(\textbf{t}^n|\alpha_l;\tau)\rho^{(l-1)}\left(\tau|\boldsymbol{\alpha}^{l-1}\right).\nonumber
		\end{align}
	Integrating this relation with respect to $\textbf{t}^n$ in the time-ordering domain $T_n$, we get for the $Q$ symbols of the POVM 
		 \begin{align}\label{Eq:POVM_Gen}
			 \Lambda^{(l)}_n(\boldsymbol{\alpha}^l)&=\Pi_n(\alpha_l)\Lambda^{(l-1)}_0\left(\boldsymbol{\alpha}^{l-1}\right)\\
			 &+\int\limits_{0}^{\tau_\mathrm{m}}d \tau \Pi_n(\alpha_l;\tau)\rho^{(l-1)}\left(\tau|\boldsymbol{\alpha}^{l-1}\right).\nonumber
		 \end{align}
	This equation connects the POVM $\hat{\Lambda}^{(l)}_n$ for the $l$th MTW in the scenario of continuous-wave detection with the POVM $\hat{\Pi}_n$ in the scenario of independent MTW and the conditional POVM $\hat{\Pi}_n(\tau)$.
	
	Let us now derive a technique for obtaining the functions $\rho^{(l-1)}\left(\tau|\boldsymbol{\alpha}^{l-1}\right)$ and $\Lambda^{(l-1)}_0\left(\boldsymbol{\alpha}^{l-1}\right)$.
	For this purpose, we explicitly separate the $n$th time in Eqs.~(\ref{Eq:POVM_Dens_Gen}) and (\ref{Eq:POVM_Gen}) as $\textbf{t}^n=(\textbf{t}^{n-1},t_n)$. 
	The function $\rho^{(k)}\left(\tau|\boldsymbol{\alpha}^{k}\right)$ can be expressed via the probability density $\lambda^{(k)}_n(\textbf{t}^{n-1},t_n|\boldsymbol{\alpha}^k)$ as
		\begin{align}\label{Eq:Rho}
			&\rho^{(k)}\left(\tau|\boldsymbol{\alpha}^{k}\right)\\
			&=\sum\limits_{n=1}^{+\infty}\int_{T_\tau^{n-1}}d^{n-1}\textbf{t}^{n-1}
			\lambda^{(k)}_n(\textbf{t}^{n-1},\tau_\mathrm{m}-\tau|\boldsymbol{\alpha}^k),\nonumber
		\end{align}
	where the term with $n=1$ assumes no integration and the integration domain $T_\tau^{n-1}$ is defined as $0\leq t_1\leq t_2\leq\ldots t_{n-1}\leq \tau_\mathrm{m}-\tau$.
	Indeed, integration with the first $(n-1)$ components of $\textbf{t}^n$ gives the probability density for the time $t_n=\tau_\mathrm{m}-\tau$ in the case of $n$ pulses and sum is taken over all $n\neq 0$. 

	Employing Eq.~(\ref{Eq:Rho}) to Eq.~(\ref{Eq:POVM_Dens_Gen}) and setting $n=0$ in Eq.~(\ref{Eq:POVM_Gen}), we arrive at a system of recurrence relations for $\rho^{(k)}\left(\tau|\boldsymbol{\alpha}^{k}\right)$ and $\Lambda_0\left(\boldsymbol{\alpha}^{k}\right)$,
		\begin{align}\label{Eq:RecRho}
			\rho^{(k)}\left(\tau|\boldsymbol{\alpha}^{k}\right)&=G(\tau|\alpha_k)\Lambda^{(k-1)}_0\left(\boldsymbol{\alpha}^{k-1}\right)\\
			&+\int\limits_{0}^{\tau_\mathrm{m}}d \tau^\prime H(\tau|\alpha_k;\tau^\prime)\rho^{(k-1)}\left(\tau^\prime|\boldsymbol{\alpha}^{k-1}\right).\nonumber
		\end{align}
		\begin{align}\label{Eq:RecLambda0}
			\Lambda^{(k)}_0(\boldsymbol{\alpha}^k)&=\Pi_0(\alpha_k)\Lambda^{(k-1)}_0\left(\boldsymbol{\alpha}^{k-1}\right)\\
			&+\int\limits_{0}^{\tau_\mathrm{m}}d \tau \Pi_0(\alpha_k;\tau)\rho^{(k-1)}\left(\tau|\boldsymbol{\alpha}^{k-1}\right).\nonumber
		\end{align}
	Here we use the notations
		\begin{align}\label{Eq:G}
			&G(\tau|\alpha_k)\\
			&=\sum\limits_{n=1}^{+\infty}\int_{T_\tau^{n-1}}d^{n-1}\textbf{t}^{n-1}\pi_n(\textbf{t}^{n-1},\tau_\mathrm{m}-\tau|\alpha_k),\nonumber
		\end{align}
	and
		\begin{align}\label{Eq:H}
			&H(\tau|\alpha_k;\tau^\prime)\\
			&=\sum\limits_{n=1}^{+\infty}\int_{T_\tau^{n-1}}d^{n-1}\textbf{t}^{n-1}\pi_n(\textbf{t}^{n-1},\tau_\mathrm{m}-\tau|\alpha_k;\tau^\prime).
			\nonumber
		\end{align}
	Resolving the recurrence relations (\ref{Eq:RecRho}) and (\ref{Eq:RecLambda0}) with the initial conditions
		\begin{align}
			\rho^{(0)}=0 \textrm{ and }\Lambda^{(0)}_0=1
		\end{align}
	we may get explicit expressions for $\rho^{(l-1)}\left(\tau|\boldsymbol{\alpha}^{l-1}\right)$ and $\Lambda^{(l-1)}_0\left(\boldsymbol{\alpha}^{l-1}\right)$.
	
	As a summary of this part we note that the POVM $\hat{\Lambda}_n$ in the photocounting formula (\ref{Eq:PhCountEq_Full_P}) and (\ref{Eq:PhotocountingEquationGeneral}) is given by Eq.~(\ref{Eq:POVM_Gen}).
	Constituents of this expression can be obtained as a solution to the system of recurrence relations given by Eqs.~(\ref{Eq:RecRho}) and (\ref{Eq:RecLambda0}).
	The POVM element $\Lambda^{(l-1)}_0\left(\boldsymbol{\alpha}^{l-1}\right)$ can in principle be excluded from Eq.~(\ref{Eq:POVM_Gen}) by employing the normalization condition (\ref{Eq:Normalization}).
	One can also formulate the recurrence relation solely for the function $\rho^{(k)}\left(\tau|\boldsymbol{\alpha}^{k}\right)$ by excluding the POVM element $\Lambda^{(k)}_0\left(\boldsymbol{\alpha}^{l-1}\right)$ in Eq.~(\ref{Eq:RecRho}).
	Similarly to the case of independent MTWs, the detection efficiency $\eta$ and dark-count intensity $\nu$ can be easily included by applying the rules (\ref{Eq:ReplRealOp}) and (\ref{Eq:ReplRealSymb}) to all coherent amplitudes or the photon-number operators.

\subsection{Approximation by the uniform distribution}

	In real practical applications, resolving the system of recurrence relations (\ref{Eq:RecRho}) and (\ref{Eq:RecLambda0}) may be an involved numerical task.
	In this subsection we introduce a reasonable approximation for this problem.
	The main condition for its applicability is stronger than the condition (\ref{Eq:CondXiMA}) for the Markovian approximation.  
	Namely, we assume that there exists a time parameter $\Delta\ll\tau_\mathrm{m}$ such that
		\begin{align}\label{Eq:CondXiUD}
			\xi(t\geq \Delta)\approx 1.
		\end{align}
	This yields 
		\begin{align}
			&\pi_n\left(\mathbf{t}|\alpha;\tau>\Delta\right)\approx\pi_n\left(\mathbf{t}|\alpha\right),\\
			&\Pi_n\left(\alpha;\tau>\Delta\right)\approx\Pi_n\left(\alpha\right).\label{Eq:POVM_Simpl}
		\end{align}
	Particularly, in the case of $\tau_\mathrm{r}=0$ the parameter $\Delta$ should be chosen as $\Delta=\tau_\mathrm{d}$. 
	Another condition for the discussed approximation assumes that the number of time bins free from dead- and relaxation-time intervals must significantly exceed $\Delta$, i.e. $\Delta\ll\tau_\mathrm{m}-|\alpha|^2(\tau_\mathrm{d}+\tau_\mathrm{r})$.
	This implies that in the domain $\tau\in[0,\Delta]$ the probability density $\rho^{(k)}\left(\tau|\boldsymbol{\alpha}^{k}\right)$ can be modeled by the uniform distribution,
		\begin{align}\label{Eq:RhoUD}
			\rho^{(k)}\left(\tau|\boldsymbol{\alpha}^{k}\right)=\frac{1-Q^{(k)}(\boldsymbol{\alpha}^{k})}{\Delta}.
		\end{align}
	Here $Q^{(k)}(\boldsymbol{\alpha}^{k})$ is the probability to get no pulses inside the time interval $\tau\in[0,\Delta]$ for the $k$th MTW. 

	In the approximation by the uniform distribution, the POVM (\ref{Eq:POVM_Gen}) is significantly simplified. 
	Firstly, we split the integration domain in the right-hand side of this equation into two parts: The first and the second ones are $[0,\Delta]$ and $[\Delta,\tau_\mathrm{m}]$, respectively.
	Next, we apply the simplification (\ref{Eq:POVM_Simpl}) to the second part.
	We also use the fact that
		\begin{align}
			\int\limits_{\Delta}^{\tau_\mathrm{m}}d \tau \rho^{(k)}\left(\tau|\boldsymbol{\alpha}^{k}\right)=Q^{(k)}\left(\boldsymbol{\alpha}^{k}\right)-\Lambda^{(k)}_0\left(\boldsymbol{\alpha}^{k}\right),
		\end{align}
	which directly follows from the normalization condition (\ref{Eq:Normalization}) and from the expression 
		\begin{align}
			1-Q^{(k)}\left(\boldsymbol{\alpha}^{k}\right)=\int\limits_{0}^{\Delta}d \tau \rho^{(k)}\left(\tau|\boldsymbol{\alpha}^{k}\right)
		\end{align}
	for the probability to get the last pulse in the domain $\tau\in[0,\Delta]$. 
	Finally, we apply the approximation (\ref{Eq:RhoUD}) for the probability density $\rho^{(k)}\left(\tau|\boldsymbol{\alpha}^{k}\right)$.
	As a result, the POVM (\ref{Eq:POVM_Gen}) is reduced to the form
		 \begin{align}\label{Eq:POVM_UD}
			\Lambda^{(l)}_n(\boldsymbol{\alpha}^l)&=Q^{(l-1)}\left(\boldsymbol{\alpha}^{l-1}\right)\Pi_n(\alpha_l)\\
			&+\frac{1-Q^{(l-1)}\left(\boldsymbol{\alpha}^{l-1}\right)}{\Delta}\int\limits_{0}^{\Delta}d \tau \Pi_n(\alpha_l;\tau).\nonumber
		\end{align}
	The information about the memory effect caused by events in the previous MTWs is encoded in this relation by the probability $Q^{(l-1)}\left(\boldsymbol{\alpha}^{l-1}\right)$. 

	The advantage of this approximation is that the probability density $\rho^{(k)}\left(\tau|\boldsymbol{\alpha}^{k}\right)$ is replaced by the probability $Q^{(k)}\left(\boldsymbol{\alpha}^{k}\right)$, which is a number depending on the coherent amplitudes in the previous MTWs.
	In order to find the corresponding recurrence relations, we integrate Eq.~(\ref{Eq:RecRho}) in the domain $[0,\Delta]$ and apply the technique used for derivation of Eq.~(\ref{Eq:POVM_UD}).
	Thus we get
		\begin{align}\label{Eq:RecRelQ}
			Q^{(l)}\left(\boldsymbol{\alpha}^{l}\right)=C\left(\alpha_l\right)Q^{(l-1)}\left(\boldsymbol{\alpha}^{l-1}\right)+B\left(\alpha_l\right),
		\end{align}
	where
		\begin{align}\label{Eq:B}
			B\left(\alpha_l\right)=1-\frac{1}{\Delta}\int\limits_{0}^{\Delta}d\tau \int\limits_{0}^{\Delta}d\tau^\prime H(\tau|\alpha_l;\tau^\prime)
		\end{align}		
		\begin{align}\label{Eq:C}
			C\left(\alpha_l\right)=A\left(\alpha_l\right)-B\left(\alpha_l\right),
		\end{align}
		\begin{align}\label{Eq:A}
			A\left(\alpha_l\right)=1-\int\limits_{0}^{\Delta}d\tau G(\tau|\alpha_k). 
		\end{align}
	The recurrence relation (\ref{Eq:RecRelQ}) should be resolved with the initial condition
		\begin{align}\label{Eq:InCondQ}
			Q^{(0)}=1,
		\end{align}
	which implies that the first MTW is not affected by the previous MTWs.
		
	The solution to the relation (\ref{Eq:RecRelQ}) with the initial condition (\ref{Eq:InCondQ}) reads
		\begin{align}\label{Eq:ProbabilityQ}
			\mathcal{Q}_{l-1}\big(\boldsymbol{\alpha}^{l-1}\big)=&B(\alpha_{l-1})+B(\alpha_{l-2})C(\alpha_{l-1})\nonumber\\
			&+B(\alpha_{l-3})C(\alpha_{l-2})C(\alpha_{l-1})+\ldots. 
		\end{align}
	The first term of this relation describes the effect of the previous MTW.
	The second term describes the effect of two previous MTWs, and so on.
	Therefore, the memory effect from the previous MTWs is conveniently encoded by the probability $\mathcal{Q}_{l-1}\big(\boldsymbol{\alpha}^{l-1}\big)$ in the POVM (\ref{Eq:POVM_UD}).

	An important feature of the solution (\ref{Eq:ProbabilityQ}) is that contributions of higher terms quickly vanish with growing their number.
	In many practical situations, it is sufficient to consider only a few first terms.  
	Therefore, only several previous MTWs affect the photocounting statistics in the given one.
	Hence, the considered measurement process is ergodic, i.e. the dependence on $l$ vanishes for $l\gg 1$ and statistics in different MTWs is almost the same. 
	This implies that the photocounting statistics of the $l$th MTW is equal to the statistics obtained from events sampled from the MTWs with consecutive numbers.
	This result justifies a technique of data processing widely used with the continuous-wave detection.
	Indeed, in typical experimental applications one averages events sampled from different MTWs without repeating the whole measurement procedure and averaging data from the MTWs with the same number.     

\subsection{Photocounting statistics vs photon-number statistics}

	An important consequence from the nonlinear dependence of the photocounting distribution on the density operator $\hat{\rho}$ in Eq.~(\ref{Eq:PhotocountingEquationGeneral}) is that the expression (\ref{Eq:PulseViaPhot}) connecting photon-number and photocounting statistics does not hold anymore.  
	Evidently, in the considered situation it should be replaced by
		\begin{align}\label{Eq:PulseStatVSPhotStat2}
			P_n=\sum\limits_{m_1,\ldots,m_l=0}^{+\infty}\Lambda_{n|m_{l},\ldots,m_1}\mathcal{P}_{m_l}
			\ldots \mathcal{P}_{m_1}.
		\end{align}	
	Here $P_n$ and $\mathcal{P}_{k}$ are photocounting and photon-number distributions, respectively, and
		\begin{align}
			\Lambda_{n|m_{l},\ldots,m_1}=\bra{m_{l},\ldots,m_1}\hat{\Lambda}_n\ket{m_{l},\ldots,m_1}
		\end{align}
	is the POVM (\ref{Eq:POVM_UD}) in the Fock representation, where we have omitted the MTW-number $l$ for the sake of simplicity. 

	Converting the POVM (\ref{Eq:POVM_UD}) in the Fock representation, we get 
		\begin{align}\label{Eq:CondProb_CW}
			&\Lambda_{n|m_{l},\ldots,m_1}\\
			&=\left[P_{n|m_l}-{D}_{n|m_l}\right]Q_{m_{l-1}\ldots m_1}+{D}_{n|m_l}.\nonumber
		\end{align}		
	Here we have introduced operators in the Fock representation. 
	The probability $P_{n|m}$ to get $n$ pulses given $m$ photons in the current MTW and no pulses in the time-interval $\tau\in[0,\Delta]$ of the previous MTW is presented by Eq.~(\ref{Eq:POVM-Fock}).
	The conditional probability ${D}_{n|m}=\bra{m}\hat{D}_n\ket{m}$ describes the situation with the presence of last pulse in the time-interval $\tau\in[0,\Delta]$ of the previous MTW.
	This probability is uniformly averaged with the time $\tau$ and corresponds to the operator
		\begin{align}\label{Eq:T}
			\hat{D}_n=\frac{1}{\Delta}\int\limits_{0}^{\Delta}d\tau\hat{\Pi}_n\left(\tau\right)
		\end{align}
   [cf. Eq.~(\ref{POVM_Symb_Tau}) for the conditional POVM $\hat{\Pi}_n\left(\tau\right)$].
	The symbol $Q_{m_{l-1}\ldots m_1}$ is decomposed according to Eq.~(\ref{Eq:ProbabilityQ}) as
		\begin{align}
			Q_{m_{l-1}\ldots m_1}=B_{m_{l-1}}&+B_{m_{l-2}}C_{m_{l-1}}\nonumber\\
			&+B_{m_{l-3}}C_{m_{l-2}}C_{m_{l-1}}+\ldots.
		\end{align}
	Here $B_m=\bra{m}\hat{B}\ket{m}$ and $C_m=\bra{m}\hat{C}\ket{m}$ correspond to the operators with the $Q$  symbols given by Eq.~(\ref{Eq:B}) and Eq.~(\ref{Eq:C}), respectively.

	The conditional probabilities $D_{n|m}$ can be obtained from Eqs.~(\ref{POVM_Symb_Tau}), (\ref{POVM_Symb_Tau_Dens}), and (\ref{Eq:T}) by applying the technique described in Sec.~\ref{Sec:Preliminaries}.
	This leads to the expression
		\begin{align}\label{Eq:POVM-FockTau}
			&D_{n|m}=\frac{m!}{\Delta(m-n)!}\\
			&\times\int\limits_{0}^{\Delta}d\tau\int_{T_n}d^n\!\mathbf{t}\, \mathcal{I}_n\left(\mathbf{t};\tau\right)\left[1-\Xi_n\left(\mathbf{t};\tau\right)\right]^{m-n}\nonumber
		\end{align}
			for $m\geq n$ and 
			\begin{align}\label{Eq:POVM-FockTau0}
			D_{n|m}=0
		\end{align}
	for $m< n$.
	The coefficients $A_m$, $B_m$, and $C_m$ are expressed via the operators $\hat{G}\left(\tau\right)$ and $\hat{H}\left(\tau|\tau^\prime\right)$ [cf. Eqs.~(\ref{Eq:B}), (\ref{Eq:C}), and (\ref{Eq:A})] in the Fock representation as
		\begin{align}
			A_m=1-\int\limits_{0}^{\Delta}d\tau G_m\left(\tau\right),
		\end{align}
		\begin{align}
			B_m=1-\frac{1}{\Delta}\int\limits_{0}^{\Delta}d\tau \int\limits_{0}^{\Delta}d\tau^\prime H_m\left(\tau|\tau^\prime\right),
		\end{align}
		\begin{align}
			C_m=A_m-B_m.
		\end{align}
	Here the functions
		\begin{align}
			G_m\left(\tau\right)&=\sum\limits_{n=1}^{m}\frac{m!}{(m-n)!}\\
			&\times\int_{T_\tau^{n-1}}d^{n-1}\textbf{t}^{n-1}
			\mathcal{I}_n\left(\textbf{t}^{n-1},\tau_\mathrm{m}-\tau\right)\nonumber\\
			&\times\left[1-\Xi_n\left(\textbf{t}^{n-1},\tau_\mathrm{m}-\tau\right)\right]^{m-n}\nonumber
		\end{align}
	and		
		\begin{align}
			H_m&\left(\tau|\tau^\prime\right)=\sum\limits_{n=1}^{m}\frac{m!}{(m-n)!}\\
			&\times\int_{T_\tau^{n-1}}d^{n-1}\textbf{t}^{n-1}
			\mathcal{I}_n\left(\textbf{t}^{n-1},\tau_\mathrm{m}-\tau;\tau^\prime\right)\nonumber\\
			&\times\left[1-\Xi_n\left(\textbf{t}^{n-1},\tau_\mathrm{m}-\tau;\tau^\prime\right)\right]^{m-n}\nonumber
		\end{align}
	are obtained from Eqs.~(\ref{Eq:G}) and (\ref{Eq:H}) with the technique described in Sec.~\ref{Sec:Preliminaries}.

\section{Examples for typical quantum states}
\label{Sec:Examples}

	In this section, we consider applications of the obtained theory for the SNSPDs to derivation of photocounting statistics for typical quantum states.
	We will concentrate on understanding roles of two factors: (1) the relaxation time $\tau_\mathrm{r}$ in the scenario of independent MTWs and (2) the memory effect of the previous MTWs in the scenario of continuous-wave detection.
	For this purpose, we compare the statistic obtained from the PNR detectors with three different situations.
	Firstly, we model the relaxation in the scenario of independent MTWs by replacing the dead time $\tau_\mathrm{d}$ with $\tau_\mathrm{d}+\tau_\mathrm{r}$ in the POVM (\ref{Eq:POVM_simple0}), (\ref{Eq:POVM_simple}), (\ref{Eq:POVM_simpleNp1}). 
	Next, we directly apply the derived POVM (\ref{POVM_NM}) with the model of smooth relaxation (\ref{Eq:RDE}).
	Finally, we consider the memory effect of the previous MTWs; cf. the POVM (\ref{Eq:POVM_UD}) and the photocounting formula (\ref{Eq:PhCountEq_Full_P}).  
	For all cases we use a monochromatic mode described by Eq.~(\ref{Eq:IntMonochr}).   
	
\begin{figure}[ht!]
		\centerline{\includegraphics[width = 1\linewidth]{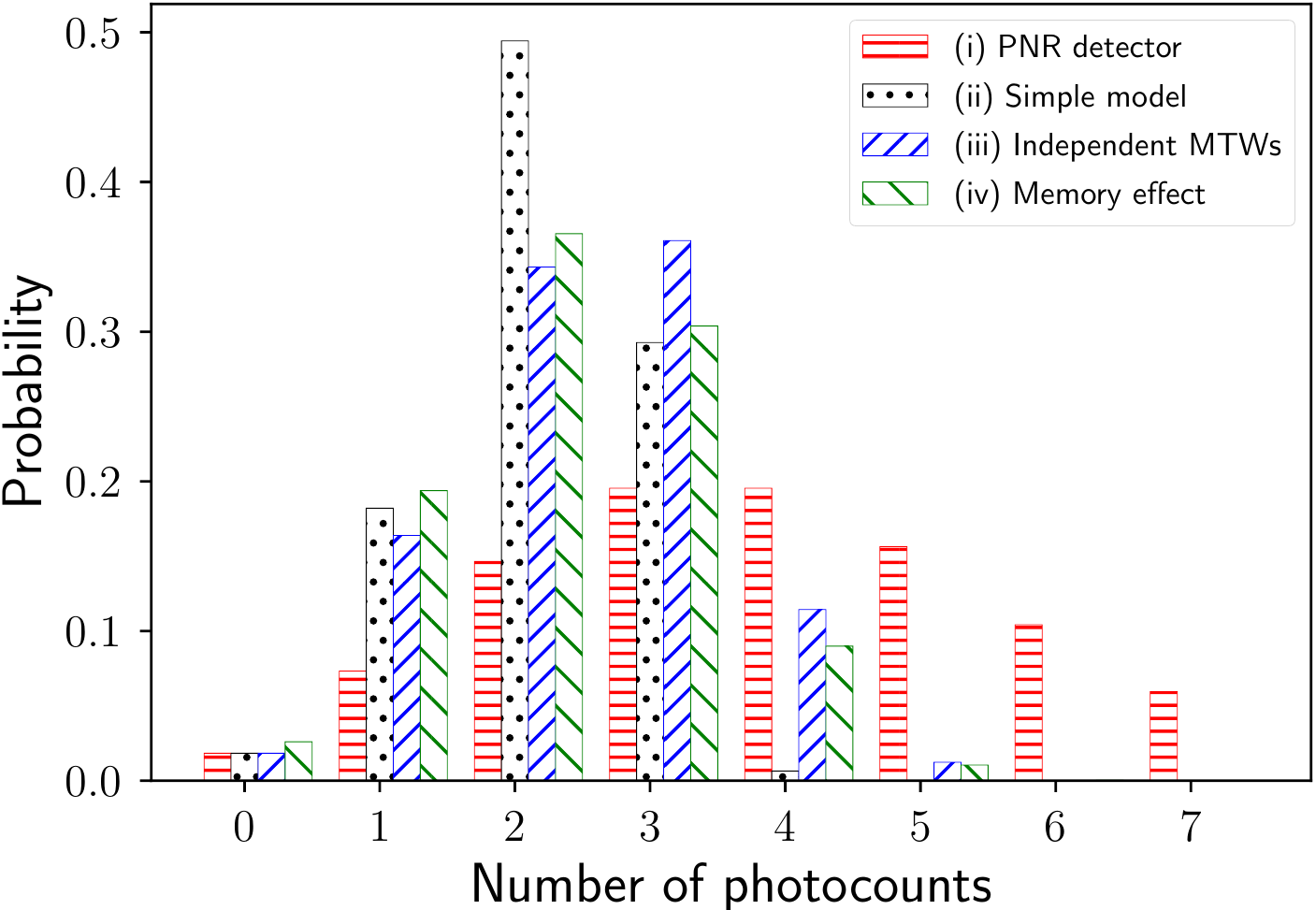}}
		\caption{\label{Fig:CS} Photocounting statistics for the coherent state $\ket{\alpha_0}$ with $\alpha_0= 2$ is shown.
			Four cases are presented: (i) PNR detectors with the statistics determined from the POVM (\ref{Eq:POVM_PNR}); (ii) the simple model described by the POVM (\ref{Eq:POVM_simple0}), (\ref{Eq:POVM_simple}), (\ref{Eq:POVM_simpleNp1}) with the dead time $\tau_\mathrm{d}$  replaced by $\tau_\mathrm{d}+\tau_\mathrm{r}$; (iii) the model given by the POVM (\ref{POVM_NM}) in the scenario of independent MTWs; (iv) the model of continuous-wave detection accounting the memory effect of the previous MTWs [cf. Eq.~(\ref{Eq:POVM_UD})] for $l=3$ and $\Delta=0.3\tau_\mathrm{m}$.
			In all appropriate cases we chose $\tau_\mathrm{d}=0.05\tau_\mathrm{m}$ and $\tau_\mathrm{r}=0.2\tau_\mathrm{m}$.}
	\end{figure}

	The first example is the coherent state $\ket{\alpha_0}$.
	Since its $P$ function is given by the Dirac delta-function,
	\begin{align}\label{Eq:CS}
	P(\alpha)=\delta(\alpha-\alpha_0),
	\end{align}
	the corresponding statistics is obtained by replacing $\alpha$ or $\alpha_k$ for all $k$ with $\alpha_0$ in the $Q$ symbols of the POVM for the scenario of independent MTWs or in continuous-wave detection, respectively.
	The result is shown in Fig.~\ref{Fig:CS}.
	It is clear that the presence of the dead time alone strictly changes the statistics in comparison with the case of the PNR detectors.
	The smooth relaxation described by Eq.~(\ref{Eq:RDE}) and the memory effect of the previous MTWs result in significant modifications of the statistics as well.  
	
\begin{figure}[htb]
		\includegraphics[width = 1\linewidth]{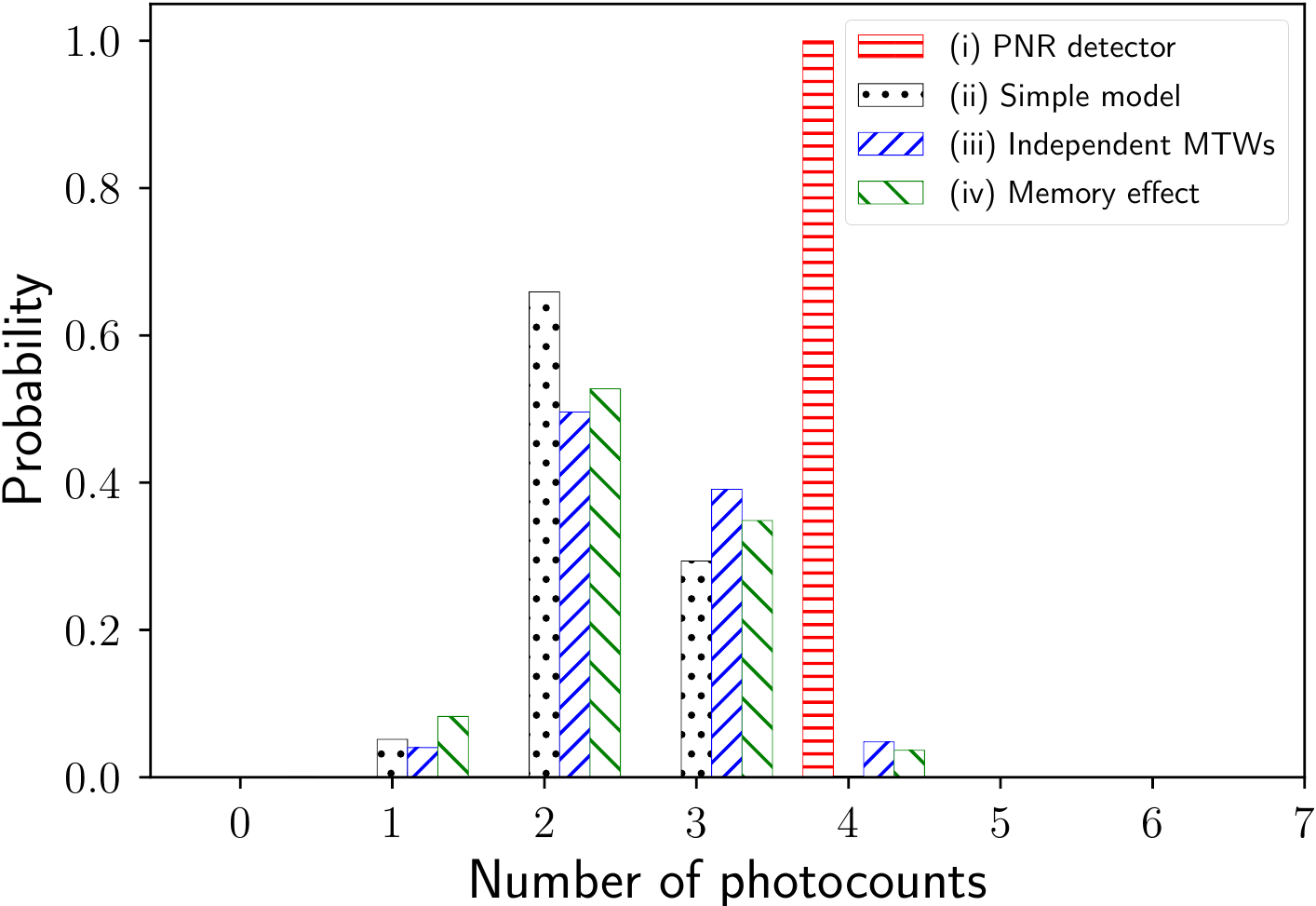}\\
		\includegraphics[width = 1\linewidth]{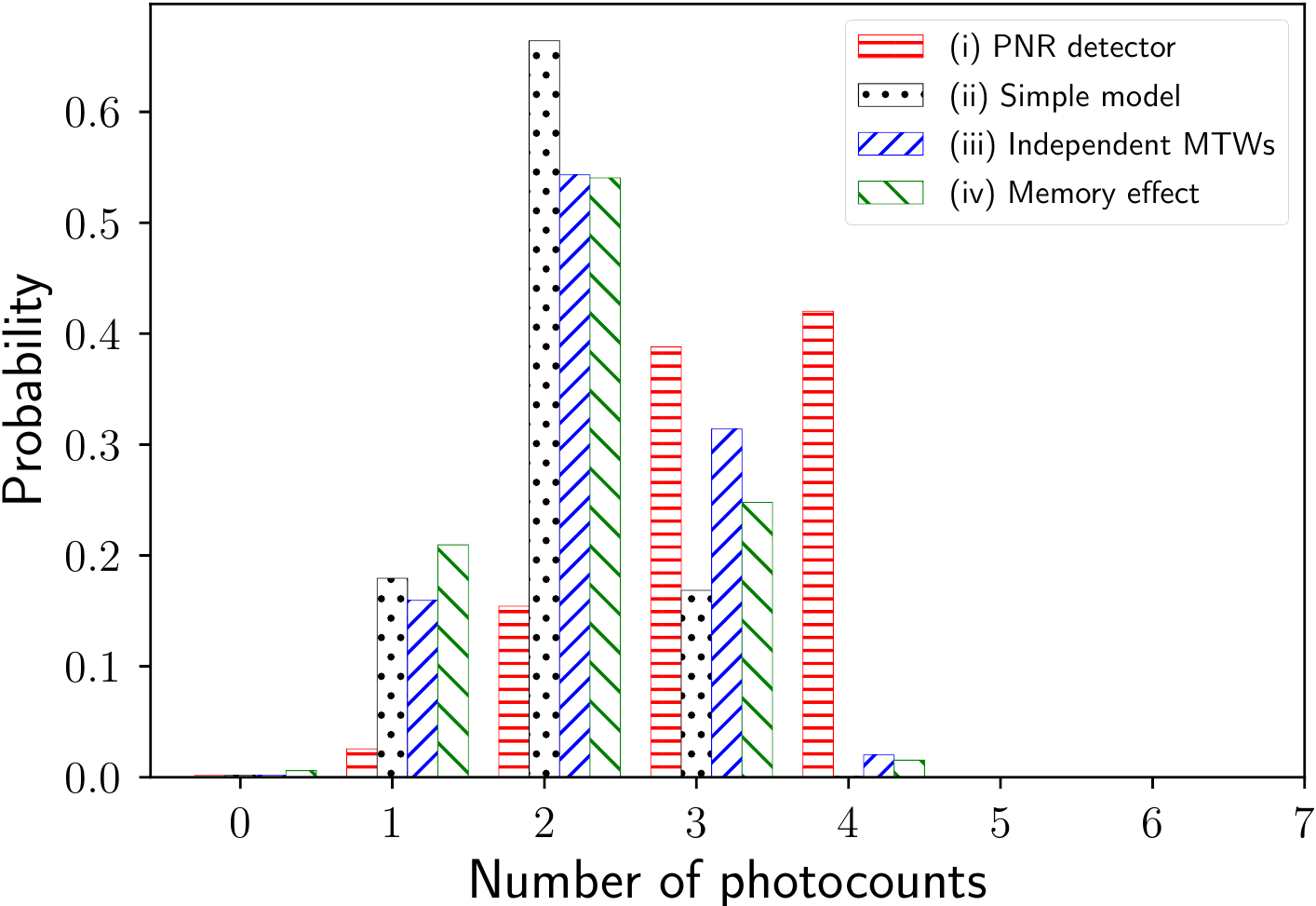}
		\caption{\label{Fig:FS}Photocounting statistics as in Fig.~\ref{Fig:CS} is shown but for the Fock state $\ket{4}$. The upper and lower plots correspond to $\eta=1$ and $\eta=0.8$, respectively.}
	\end{figure}

	As the next example, we consider the Fock state $\ket{k}$ attenuated with the efficiency $\eta$, which can also be considered as the detection efficiency.
	The corresponding density operator is given by
	\begin{align}\label{Eq:Fock}
	\hat{\rho}=\sum\limits_{l=0}^{k}\mathcal{P}_l\ket{l}\bra{l},
	\end{align}
	where
	\begin{align}\label{Eq:FockPND}
	\mathcal{P}_l=\binom{k}{l}\eta^l(1-\eta)^{k-l}
	\end{align}	
	is the photon-number distribution.	
	The photocounting distribution is directly obtained from Eq.~(\ref{Eq:PulseViaPhot}) and Eq.~(\ref{Eq:PulseStatVSPhotStat2}) in the scenarios of independent MTWs and continuous-wave detection, respectively. 
	It is shown in Fig.~\ref{Fig:FS} for $k=4$.
	Firstly, it is clearly seen that even the dead-time alone results in a significant increase of events with $n<k$.
	Secondly, the consideration of a smooth model for the relaxation crucially changes the probability distribution as well. 
	Finally, we note the significant impact of the previous MTWs. 
	
\begin{figure}[h]
			\includegraphics[width = 1\linewidth]{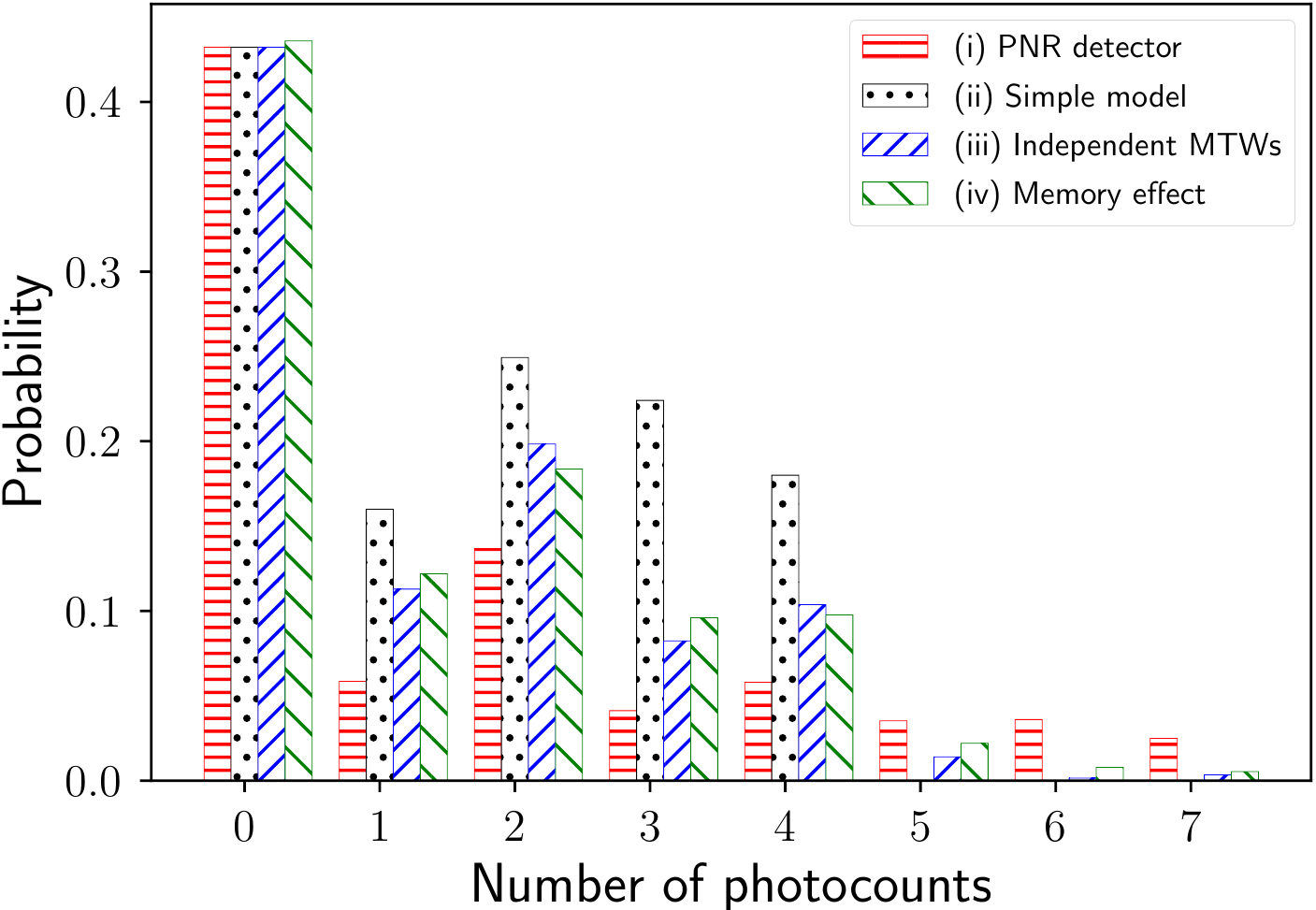}
			\caption{\label{Fig:SqS} Photocounting statistics as in Fig.~\ref{Fig:CS} is shown but for the squeezed vacuum state $\ket{r}$. The squeezing parameter and the detection efficiency are chosen as $r=1.5$ and $\eta=0.8$, respectively.}
		\end{figure}

	As the last example, we consider the squeezed vacuum state,
		\begin{align}\label{Eq:SqS}
			\ket{r}=\frac{1}{\sqrt{\cosh r}}\sum_{n=0}^{\infty}\binom{2n}{n}^{1/2}\left(\frac{\tanh r}{2}\right)^n\ket{2n},
		\end{align}
	where $r$ is the squeezing parameter.
	The corresponding photocounting distribution can be obtained with two different methods.
	The first one is based on the fact that the photon-number distribution for this state is given by 
		\begin{align}
			\mathcal{P}_n&=\frac{(i\sinh r)^n}{[1+(2-\eta)\eta\sinh^2r]^{\frac{n+1}{2}}}\\
			&\times\Legendre_n\left[\frac{(1-\eta)\eta\sinh r}{i \sqrt{1+(2-\eta)\eta\sinh^2r}}\right],
			\nonumber
		\end{align}
	where $\Legendre_n(x)$ are the Legendre polynomials.
	These probabilities can be substituted in Eqs.~(\ref{Eq:PulseViaPhot}) and (\ref{Eq:PulseStatVSPhotStat2}) in order to obtain the photocounting distribution for the SNSPDs in the scenarios of independent MTWs and continuous-wave detection, respectively.
	Alternatively, for the scenario of independent MTWs one can directly use the photocounting formula (\ref{Eq:PhotocountingEquationHS}) with the POVM (\ref{POVM_NM}) and get the photocounting distribution in the form
		\begin{align}\label{Eq:PNDvPNDens}
			P_n=\int_{T^n} d^n\textbf{t} p_n(\textbf{t}),
		\end{align}
	where
		\begin{align}
			p_n(\textbf{t})=\Tr\left[\hat{\pi}_n(\textbf{t})\hat{\rho}\right]
		\end{align}
	is the unnormalized probability density to get pulses at the time moments $\textbf{t}$, $\hat{\rho}=\ket{r}\bra{r}$ is the density operator, and $\hat{\pi}_n(\textbf{t})$ is the operator, the $Q$ symbol of which is given by $\pi_n(\textbf{t}|\alpha)$, cf. Eq.~(\ref{Eq:POVM_DT_gen}).
	The function $p_n(\textbf{t})$ can be obtained explicitly as
		\begin{align}\label{Eq:PNDens_Sq}
			p_n(\textbf{t})&=\frac{n!i^n \mathcal{I}_n(\textbf{t})\sinh^nr} {[S_n(\textbf{t})]^{n+1}}\\
			&\times\Legendre_n\left(\frac{i\sinh r[\Xi_n(\textbf{t})-1]}{ S_n(\textbf{t})}\right),
			\nonumber
		\end{align}
	where
		\begin{align}
			S_n(\textbf{t})={\sqrt{1-\sinh^2r\Xi_n(\textbf{t})\left[\Xi_n(\textbf{t})-2\right]}}.
		\end{align}
	With these expressions, the integral in Eq.~(\ref{Eq:PNDvPNDens}) can be evaluated numerically.
	Similar calculations can be directly conducted for the scenario of continuous-wave detection.
	The result is presented in Fig.~\ref{Fig:SqS}.
	Evidently, dead and relaxation times as well as the memory effect of the previous MTWs play a crucial role in the photocounting statistics for the squeezed vacuum states. 
	
\section{Reconstruction of the time-dependent efficiency}
\label{Sec:Xi}

	For our purposes, we have used the time-dependent efficiency in the form of a decaying exponent as is given by Eqs.~(\ref{Eq:TDE}) and (\ref{Eq:RDE}).
	However, the realistic function may differ from this simple model.
	In this section we analyze a technique reported in Ref.~\cite{Autebert2020}, enabling reconstruction of the efficiency $\xi(t)$ from the experimental data.
	
	Let us consider a monochromatic wave in the coherent state $\ket{\alpha_0}$ in the MTW of duration $\tau_\mathrm{m}$.
	The intensity of this wave (the mean energy per time unit) is given by the value $\lambda=|\alpha_0|^2/\tau_\mathrm{m}$. 
	We will derive the probability density for the time between pulses in such a scenario.
	It is composed of two parts:
		\begin{itemize}
			\item The probability density to get the second pulse at the time moment $t$ given the first pulse registered at the time moment $t=0$ in the form $\lambda\xi(t)$.
			\item The probability to get no pulses in the time domain $(0,t)$ given by
			\begin{align}
				\exp\left[-\lambda\int\limits_{0}^{t}d \tau \xi(\tau)\right].
			\end{align}
		\end{itemize}
	The resulting probability density for the time between pulses is given by
		\begin{align}\label{Eq:DistrBetwPuls}
			P(t)=\lambda\xi(t)\exp\left[-\lambda\int\limits_{0}^{t}d \tau \xi(\tau)\right],
		\end{align}
	which is the product of two mentioned components.
	
	Similar to Eq.~(\ref{Eq:CondXiUD}) we assume the existence of time $\Delta$, for which the relaxation processes are finished.
	Let the intensity be chosen such that the condition
		\begin{align}
			\lambda\ll \left[\int\limits_{0}^{\Delta}d \tau \xi(\tau)\right]^{-1}
		\end{align}
	is satisfied.
	For times $t\leq\Delta$ the probability density (\ref{Eq:DistrBetwPuls}) can be approximated as
		\begin{align}
			P(t)\approx\lambda\xi(t).
		\end{align}
	Therefore, the probability density of the time between pulses for $t\leq\Delta$ is proportional to the time-dependent efficiency $\xi(t)$.
	This fact can be used for its reconstruction in experiments. 
	

\section{Summary and conclusions}
\label{Sec:Concl}

	To summarize, we have proposed a photodetection theory of the SNSPDs for the technique of counting voltage pulses, which occur during the MTWs.
	We have concentrated with three issues: dead time, relaxation time, and the memory effect from the previous MTWs.
	The latter can be eliminated with the considered here technique of independent MTWs assuming either darkening detector input after each MTW or a proper postselection of the MTWs.

	Our idea is based on modeling the dead time and the relaxation process by the time-dependent efficiency.
	This efficiency becomes zero after each pulse during the dead-time interval and then it is smoothly recovered.
	For a better understanding of the role of dead and relaxation times, the recovering part of this efficiency has been approximated by the exponential function.
	However, the realistic dependence may have a different form, which can also be applied in our theory.
	We have analyzed an experimental technique of its reconstruction from experimental data.

	The measurement technique of continuous-wave detection assuming no interruptions between the MTWs is considered in the Markovian approximation.
	In this case, the well-known photodetection formula significantly differs from its standard form.
	Indeed, the memory effect of the previous MTWs results in nonlinear dependence of the photocounting distribution on the density operator.
	
	In the most general case, photocounting statistics in each MTW depends on its number.
	We have shown that in the Markovian scenario this dependence is weak such that the measurement process can be considered as ergodic.	
	This result justifies a widely-used experimental technique of averaging results from the events sampled in all MTWs.
	
	The SNSPDs are widely used in many experimental works in quantum optics.
	They also successfully applied for practical implementations of quantum-information technologies.
	We hope that our results will be useful for a proper analysis of corresponding experimental data and for further development of modern quantum technologies.

\begin{acknowledgements}
	The authors acknowledge support from the National Research Foundation of Ukraine through the project 2020.02/0111 ``Nonclassical and hybrid correlations of quantum systems under realistic conditions''.
	They also thank B. Hage and J. Kr\"oger for enlightening discussions.
\end{acknowledgements}

\bibliography{biblio}

\end{document}